\documentclass[runningheads]{llncs}
\usepackage{graphicx}
\usepackage{subfigure}
\usepackage{comment}
\usepackage[export]{adjustbox}[2011/08/13]
\usepackage{cite}
\usepackage{xcolor}
\usepackage{tabularx}
\usepackage[backref]{hyperref}
\usepackage{float}
\usepackage[tableposition=top]{caption}
\usepackage{booktabs}
\usepackage{amsmath}
\usepackage{amsfonts}
\usepackage[misc]{ifsym}

\begin{document}
%
\title{CT reconstruction from few planar X-rays with application towards low-resource radiotherapy}
%
\titlerunning{Low-resource radiotherapy planning from few planar X-rays}

\author{Yiran Sun\inst{1} \and
Tucker Netherton\inst{2} \and
Laurence Court\inst{2} \and Ashok Veeraraghavan\inst{1}\textsuperscript{(\Letter)} \and Guha Balakrishnan\inst{1}\textsuperscript{(\Letter)}}
%
\authorrunning{Y. Sun et al.}
%
\institute{Department of Electrical and Computer Engineering, Rice University\and
Department of Radiation Physics, University of Texas MD Anderson Cancer Center\\
\email{\{ys92, vashok, guha\}@rice.edu, \{TNetherton, LECourt\}@mdanderson.org}}
%
\maketitle 

%
\begin{abstract}

CT scans are the standard-of-care for many clinical ailments, and are needed for treatments like external beam radiotherapy. Unfortunately, CT scanners are rare in low and mid-resource settings due to their costs. Planar X-ray radiography units, in comparison, are far more prevalent, but can only provide limited 2D observations of the 3D anatomy. In this work, we propose a method to generate CT volumes from few ($<5$) planar X-ray observations using a prior data distribution, and perform the first evaluation of such a reconstruction algorithm for a clinical application: radiotherapy planning. We propose a deep generative model, building on advances in neural implicit representations to synthesize volumetric CT scans from few input planar X-ray images at different angles. To focus the generation task on clinically-relevant features, our model can also leverage anatomical guidance during training (via segmentation masks). We generated 2-field opposed, palliative radiotherapy plans on thoracic CTs reconstructed by our method, and found that isocenter radiation dose on reconstructed scans have $<1\%$ error with respect to the dose calculated on clinically acquired CTs using $\leq 4$ X-ray views. In addition, our method is better than recent sparse CT reconstruction baselines in terms of standard pixel and structure-level metrics (PSNR, SSIM, Dice score) on the public LIDC lung CT dataset. Code is available at: \url{https://github.com/wanderinrain/Xray2CT}.

\end{abstract}
\section{Introduction}
CT scans are the standard-of-care for diagnosis and treatment of many diseases. However, due to their costs and infrastructure requirements, global inequities in access to CT scanners exist in many low-to-middle income countries (LMICs)~\cite{hricak2021medical}. This lack of CT access impacts many facets of healthcare such as external beam radiotherapy, in which a treatment planning system calculates the ionizing dose to a patient's tumor and surrounding tissues by utilizing the electron density information from the CT voxels. In comparison, planar X-ray units are far more prevalent in LMICs than CT units, and recent studies~\cite{shen2019patient,ying2019x2ct} demonstrate that significant information in CT scans may be estimated from sparse observations using deep generative networks trained over large datasets. With this motivation, we propose a learning-based algorithm for synthesizing CT volumes from few ($<5$) planar X-ray images, and demonstrate basic feasibility for radiotherapy planning for post-mastectomy chest walls (extremely prevalent for women in low-resource settings).

State-of-the-art CT reconstruction methods from sparse views are based on learning complex priors with neural networks and operate in both the sinogram~\cite{sun2021coil, shen2022nerp}, and voxel~\cite{ying2019x2ct, ge2022x, jiang20203d, shen2019patient} spaces. Several voxel-based methods use convolutional neural networks (CNNs) optimized on (CT, X-ray) supervised pairs with $<5$ views~\cite{ying2019x2ct, shen2019patient, jiang20203d}. Others use implicit neural representations (INRs), networks that map voxel coordinates to intensity values and can better reconstruct high-frequency details than CNNs \cite{xie2022neural}. However, INRs are typically fit using only the input views (i.e., self-supervised), and so require at least 20 views to attain reasonable results~\cite{zha2022naf}. If such an approach were used with planar radiography, the large number of planar image acquisitions becomes practically infeasible, as technologists would need to reposition the patient and detector per orientation. In addition, previous studies provide limited evaluation, using only pixel-level reconstruction metrics like PSNR and SSIM~\cite{wang2004image}.

To the best of our knowledge, we propose the first supervised CT reconstruction algorithm from few ($<$ 5) planar X-ray views using INRs. We build on the \textit{pixelNeRF}~\cite{yu2021pixelnerf} model design for sparse view synthesis problems. Our model first extracts 2D feature images from each input planar X-ray using a CNN U-Net\cite{ronneberger2015u}. For each 3D coordinate, it then uses an INR to predict the output CT's intensity given the coordinate, and a set of 2D features obtained by projecting the coordinate onto each feature image based on the known geometry of the X-ray imaging system. Our training loss function includes both a typical reconstruction term, and a segmentation term (captured by a pretrained segmentation network) which we hypothesize will be useful because radiotherapy plans rely on accurate anatomical boundaries.

We evaluated our method on reconstructing CT scans from 1 to 4 input planar X-rays. First, our method outperforms neural network baselines on the public LIDC-IDRI \cite{armato2011lung} lung CT dataset in terms of pixel-level (PSNR, SSIM) and structural (Dice Similarity Coefficient (DSC)\cite{dice1945measures}) metrics. Next, we evaluated our method using an in-house thoracic CT dataset for post-mastectomy chest wall radiotherapy, in which the tumor has been removed prior to the acquisition of the CT scan and the target of the radiotherapy would need only consider organs within the CT (e.g., chest wall, lungs, heart, spinal cord). 2-field opposed radiotherapy plans generated from our model's reconstructions obtain $< 1\%$ error with respect to isocenter dose compared to clinical scans, well below the criterion for dose verification accuracy~\cite{zhu2021report}. We conclude by discussing limitations and steps to move towards clinical application.

\section{Method}

\begin{figure}[t!]
    \centering
    \includegraphics[width=\textwidth]{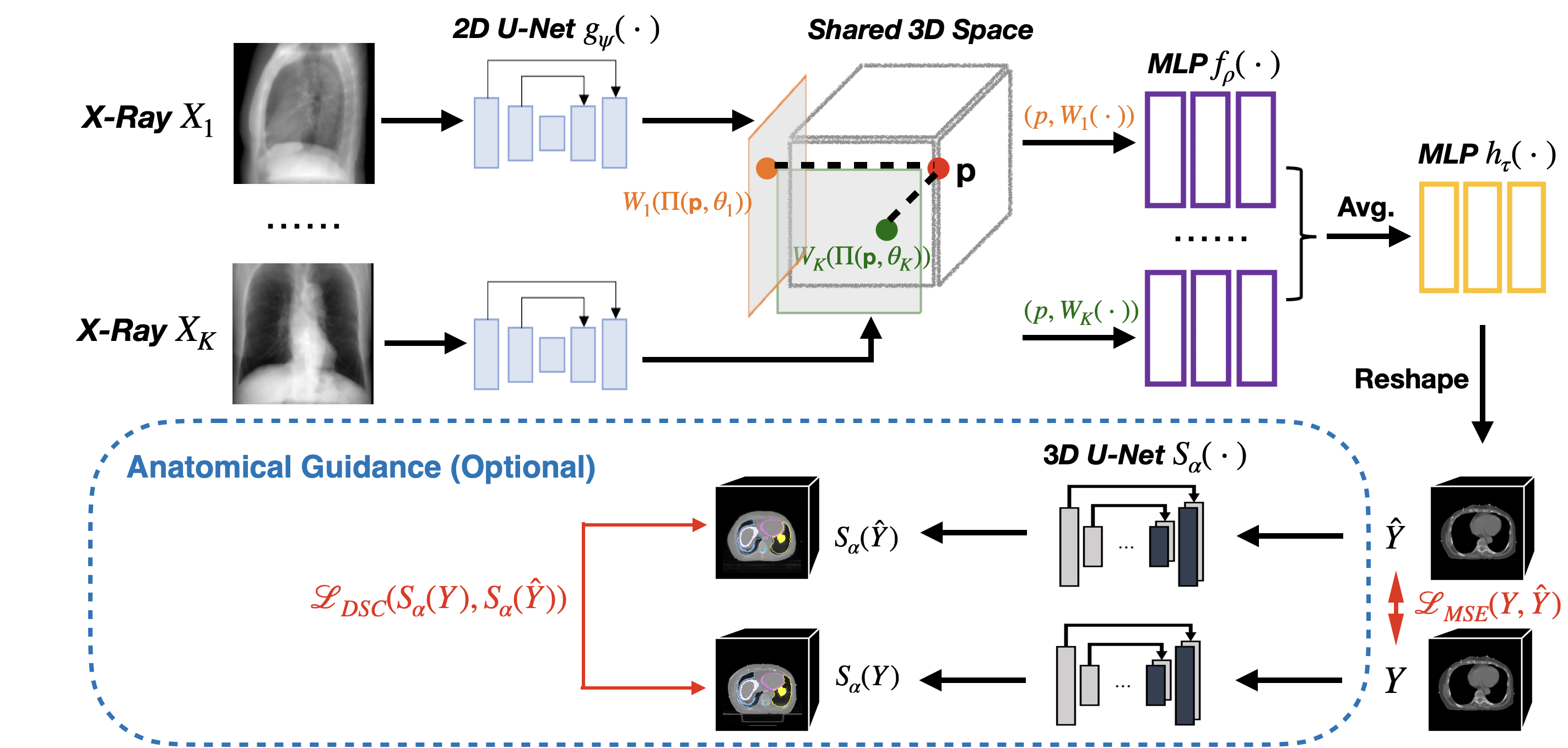}
    \caption{\textbf{Method overview.} Our model takes different planar X-ray views as input and outputs a predicted CT volume $\hat{Y}$. 2D U-Net $g_\psi$ generates a feature image $W_i$ from each $X_i$. Then, for each 3D point $\textbf{p}$, the projection operator $\Pi$ retrieves the aligned feature vector $W_i(\Pi(\textbf{p}, \theta_i))$ from each view and passes them into MLPs $f_\rho$ and $h_\tau$ to predict intensity $\hat{Y}(\textbf{p})$. We train the model with a reconstruction loss and an optional loss based on anatomical segmentation overlap (blue box).}
    \label{fig:method}
\end{figure}

Let $\textbf{\textit{X}} = \left\{X_1, \cdots, X_K \right\}$ represent $K$ input planar X-rays acquired from different orientations $\left\{\theta_1, \cdots \theta_K\right\}$, where $X_i \in \mathbb{R}^{d \times d \times 1}$, and $Y \in \mathbb{R}^{d \times d \times d \times 1}$ represents the associated ground truth CT volume\footnote{We assume a uniform dimension $d$ here for simplicity, but our method can handle arbitrary dimensions.}.
Our goal is to learn a model that maps $\textbf{X}$ to $Y$. The main challenge of this reconstruction task is to combine the information from the different X-ray views into one shared 3D space. The overview of key components in our approach is illustrated in Fig. \ref{fig:method}.

Building on \textit{pixelNeRF}\cite{yu2021pixelnerf}, we propose a model with three components: a 2D feature extraction network $g_\psi(\cdot)$ that extracts planar X-ray image features, a projection operation $\Pi(\cdot, \cdot)$ that maps 3D coordinates and a viewing angle to 2D coordinates, and an INR (implemented with functions $f_\rho(\cdot)$ and $h_\tau(\cdot)$) that maps 3D coordinates and $K$ 2D feature vectors to voxel intensities. We supervise the entire model with a loss function consisting of a reconstruction error, and an (optional) segmentation error penalizing incorrect anatomical boundaries. \\

\noindent \textbf{X-ray CNN: } We implement function $g_\psi(\cdot)$ with a 2D CNN U-Net~\cite{ronneberger2015u}, which outputs an image $W_i \in \mathbb{R}^{d \times d \times c}$ encoding $c$ multiscale features per pixel for $X_i$. \\

\noindent \textbf{Projection: }
$\Pi(\cdot, \cdot): \mathbb{R}^{3} \times \mathbb{R}^{1} \rightarrow \mathbb{R}^{2}$ maps a 3D coordinate and angle $\theta_i$ to the corresponding 2D location on image $W_i$ based on the known X-ray imager geometry. For example, if the X-rays were generated via parallel beam radiation, each point will be orthogonally projected onto $W_i$ along angle $\theta_i$. For fan-beam radiation, each point will be projected based on rays emanating from a 3D source point. 
The output of this operator is a feature vector $W_i(\Pi(\textbf{p}, \theta_i)) \in \mathbb{R}^{c}$. \\

\noindent \textbf{Conditional INR: } Next, we use features $\left\{W_i(\Pi(\textbf{p}, \theta_i))\right\}_{i=1}^K$ to estimate the voxel intensity at location $\textbf{p}$. We use two multilayer perceptrons (MLPs) to do this: $f_\rho (\cdot, \cdot)$ and $h_\tau (\cdot)$. $f_\rho \in \mathbb{R}^2 \times \mathbb{R}^c \rightarrow \mathbb{R}^h$ operates on each view independently, and is responsible for combining a Fourier feature transform~\cite{tancik2020fourier} of $\textbf{p}$ and feature vector $W_i(\Pi(\textbf{p}, \theta_i)))$ from view $i$ into an embedding $\textbf{r}_i(\textbf{p})$. Fourier feature coordinate transforms empirically result in better high-frequency reconstructions compared to the coordinates on their own. Next, we compute the average embedding over all views $\hat{\textbf{r}}(\textbf{p})$, 
and feed it into MLP $h_\tau (\cdot)$, which outputs $\hat{Y}(\textbf{p})$, an estimate of the intensity value (a scalar) at $\textbf{p}$. We use three residual blocks for both MLPs, containing fully-connected linear layers with 128 neurons and sinusoidal periodic activation functions \cite{sitzmann2020implicit}. \\

\noindent \textbf{Loss Function: } We train our model using the loss: $\mathcal{L}_{total} = \| \hat{Y}-Y\|_2^2 + \lambda \cdot \mathcal{L}_{DSC}(S_\alpha(\hat{Y}), S_\alpha(\hat{Y}))$, consisting of a typical mean squared error (MSE) term, and an (optional) term evaluating Dice score~\cite{dice1945measures} between the segmentation masks of the two scans, estimated by pretrained segmentation network $S_\alpha(\cdot)$.

\section{Experiments}

\begin{figure}[t!]
    \centering
    \includegraphics[width=\textwidth, center]{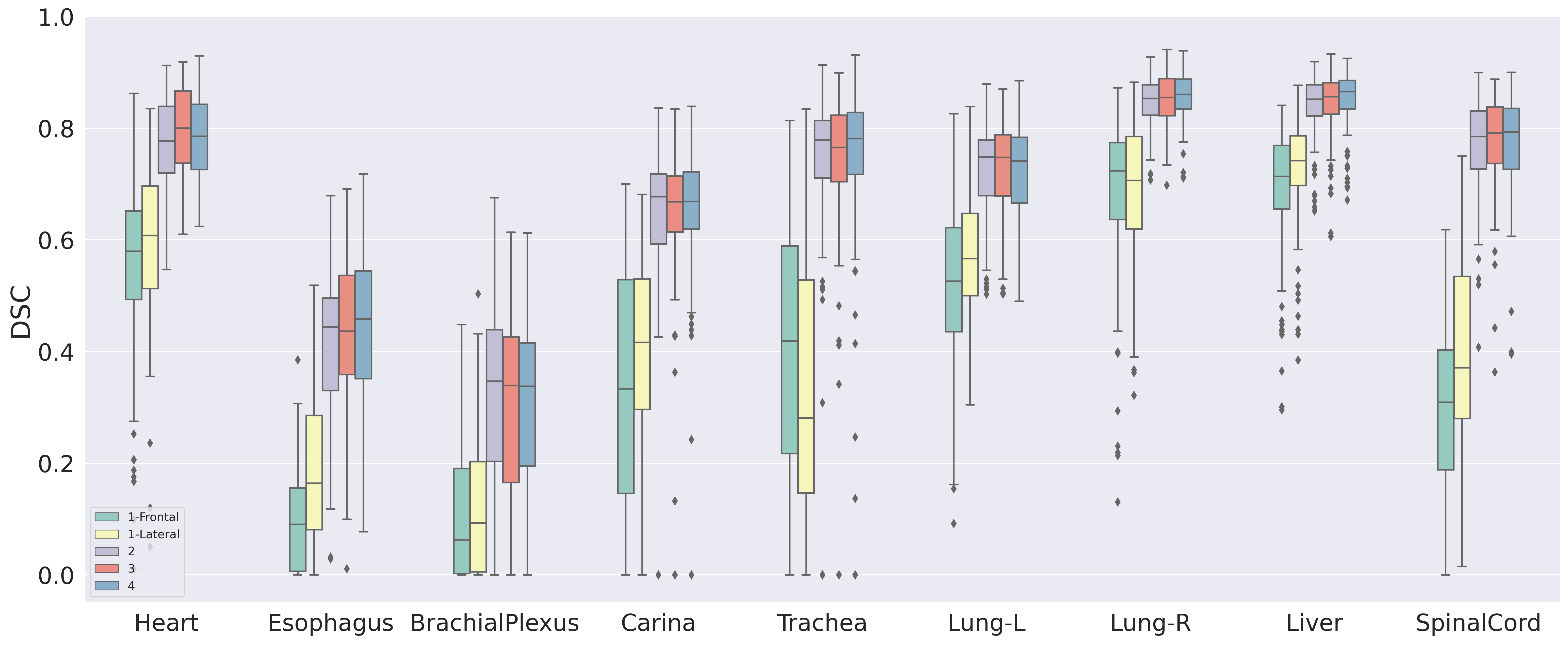}
    \caption{\textbf{Boxplot of Dice similarity coefficients (DSC) between reconstructed and ground truth CT scans using the Thoracic dataset.} We compare versions of the proposed model Ours-Seg with different numbers of input views, on 100 test subjects. Higher values are better.}
    \label{fig:breast-dsc}
\end{figure}

We evaluated our model using the public Lung Image Database Consortium (LIDC-IDRI)~\cite{armato2011lung} lung CT dataset, and an in-house Thoracic CT dataset from patients who received radiotherapy (gathered under an IRB approved protocol). LIDC includes 1018 patients, which we randomly split into 868/50/100 train/validation/test groups, and Thoracic includes 997 patients which we randomly split into 850/47/100 train/validation/test scans. We clipped all voxel values to $[-1000, 1000]$ Hounsfield Units (HU). We resampled each scan to $1$ mm$^3$ resolution, cropped it to a cube, and then resized it to $128^3$ voxels. We generated four planar X-ray views per CT at angles of: $0^{\circ}$ (Lateral), $45^{\circ}$, $90^{\circ}$ (Frontal), and $135^{\circ}$ using the Digitally Reconstructed Radiograph (DRR) generator \textit{Plastimatch}\cite{sharp2010plastimatch}, with energy level 50keV. For our segmentation loss, we trained one segmentation network per dataset using a UNet~\cite{balakrishnan2019voxelmorph,ronneberger2015u}. For LIDC, we trained the segmentation network on 3 structures (left \& right lung, nodule) using the LUNA16\cite{setio2017validation} dataset. For Thoracic, we trained on 9 structures (see Fig.~\ref{fig:breast-dsc} for names) predicted by nnUNet network~\cite{isensee2021nnu} used for contouring in the clinic. In the following results, we call our models trained without the segmentation loss \emph{Ours}, and those trained with the segmentation loss \emph{Ours-Seg}.

\noindent \textbf{Metrics:} We evaluated performance using three types of metrics: voxel level (PSNR, SSIM~\cite{wang2004image}), structural level (Dice Similarity Coefficient~\cite{dice1945measures}, or DSC), and radiation dose (Isocenter dose). Isocenter dose is defined as the calculated dose (in centigray or cGy) deposited to a point in the patient's body at a distance of 100 cm away from a megavoltage X-ray source. 

\noindent \textbf{Baselines:} We experimented with two neural network baselines: X2CT-CNN~\cite{ying2019x2ct} and Neural Attenuation Fields (NAF)~\cite{zha2022naf}. X2CT-CNN is a CNN for reconstructing CT scans from 1 or 2 (orthogonal) views. NAF is a recently proposed implicit neural representation (INR) that handles arbitrary viewing angles in CT reconstruction (like our model), but uses no prior training data. We trained both baselines from scratch on each dataset separately. 

\noindent \textbf{Implementation:} We implemented our models in PyTorch~\cite{paszke2019pytorch} and ran all experiments on NVIDIA A100 GPUs with 40/80 GB of memory. We set the batch size to 1 and trained for 100 epochs per model.
We used the ADAM~\cite{kingma2014adam} optimizer with an initial learning rate of 3e$^{-5}$, and decreased the learning rate to 3e$^{-6}$ after 50 epochs. 

\noindent \textbf{Radiotherapy planning:} Using 10 randomly selected patients from the Thoracic dataset, we generated radiotherapy plans with 2-field opposed beam arrangements using the RayStation commercial treatment planning system~\cite{bodensteiner2018raystation}. We set the isocenter within the thoracic spine of each clinical CT, fractional dose to 200 cGy, beam energy to 15 MV, and the radiation field size to 10 $\times$ 10 cm$^2$ at isocenter. We performed rigid image registration so that radiation plans may be directly compared between the clinical and reconstructed CTs. Isocenter dose was compared between clinical plans and plans made on reconstructed CTs.

\begin{figure}[t!]
\subfigure{
	\begin{minipage}[t]{0.32\linewidth}
		\centering
		\includegraphics[width=1.6in]{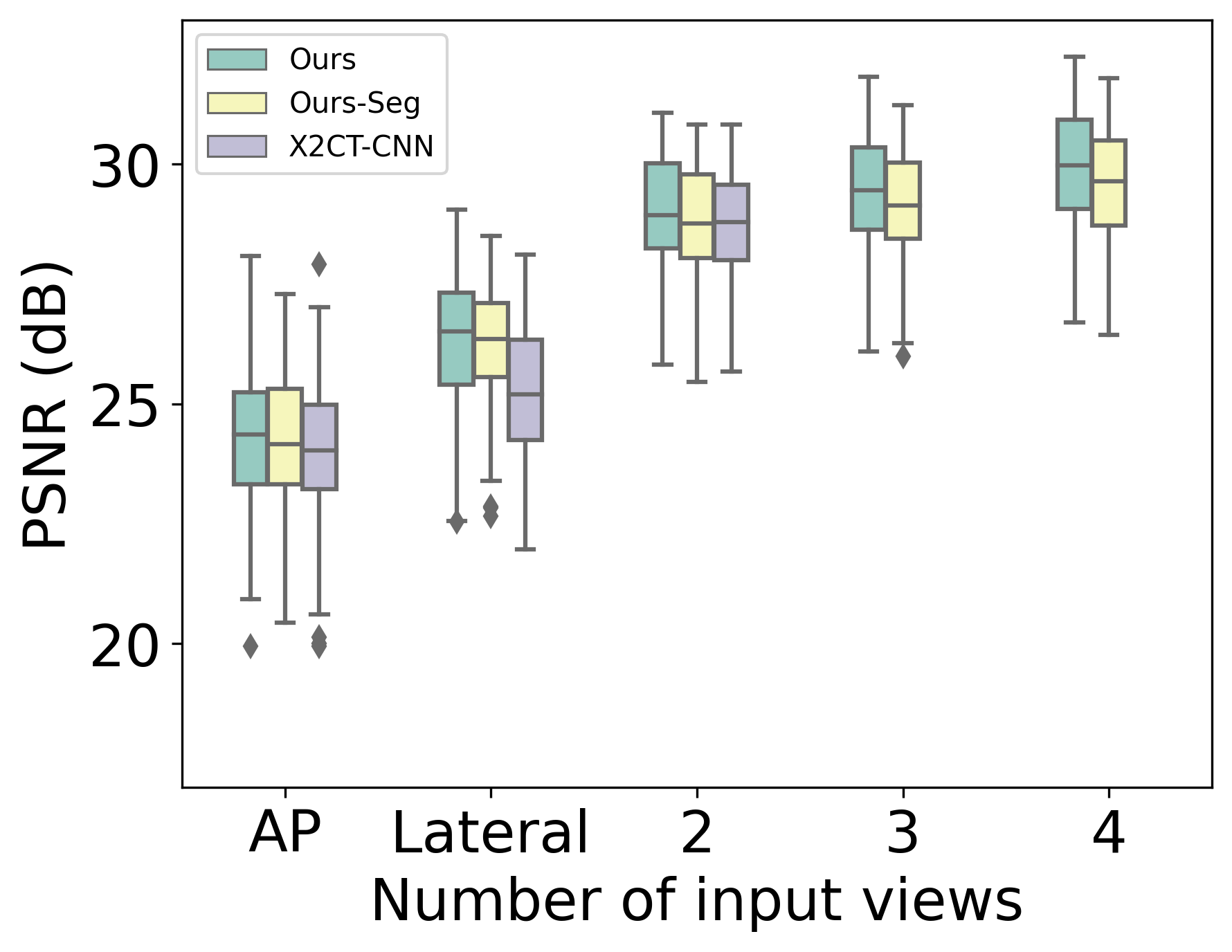}
	\end{minipage}
	\begin{minipage}[t]{0.32\linewidth}
		\centering
		\includegraphics[width=1.6in]{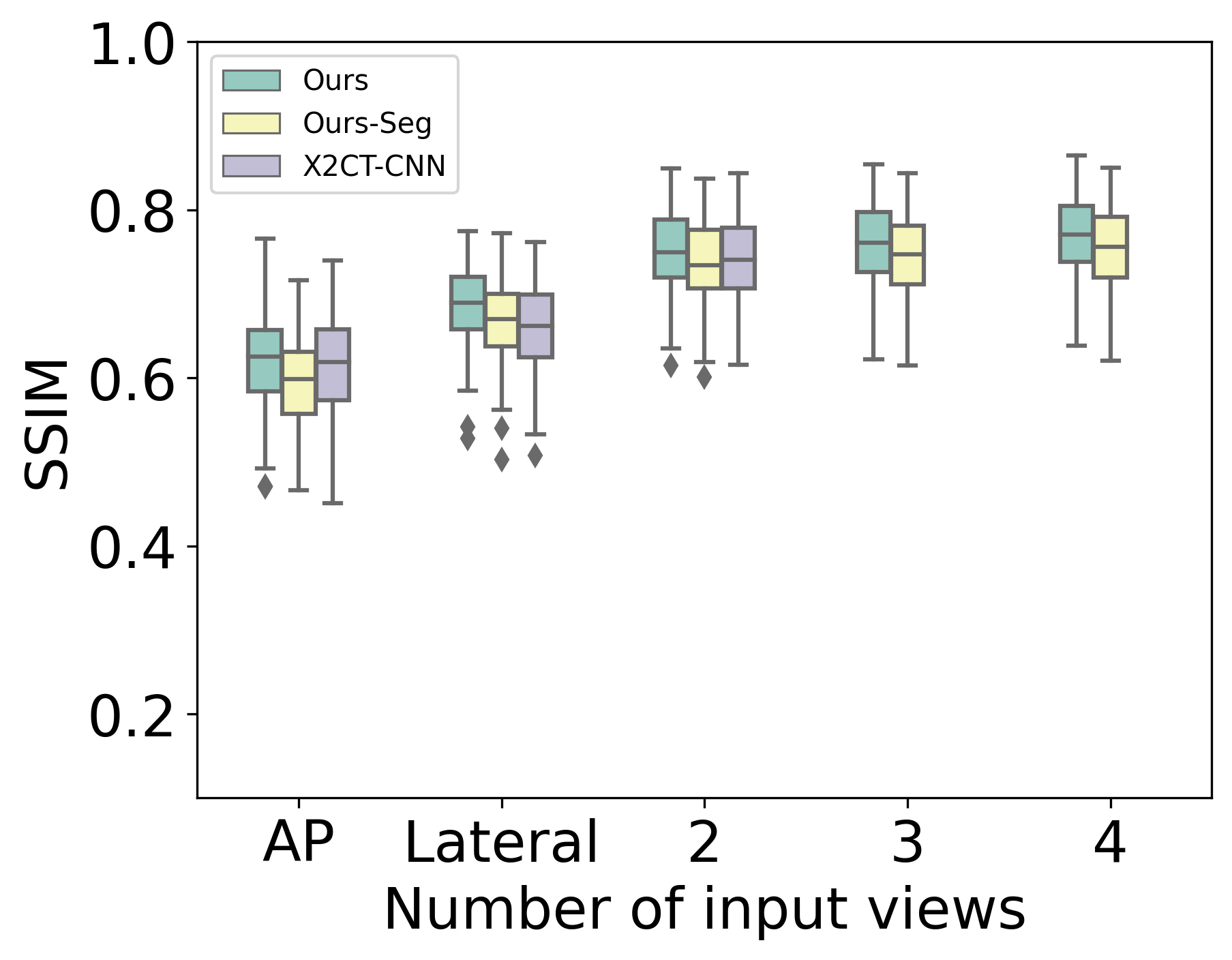}
	\end{minipage}
 	\begin{minipage}[t]{0.32\linewidth}
		\centering
		\includegraphics[width=1.6in]{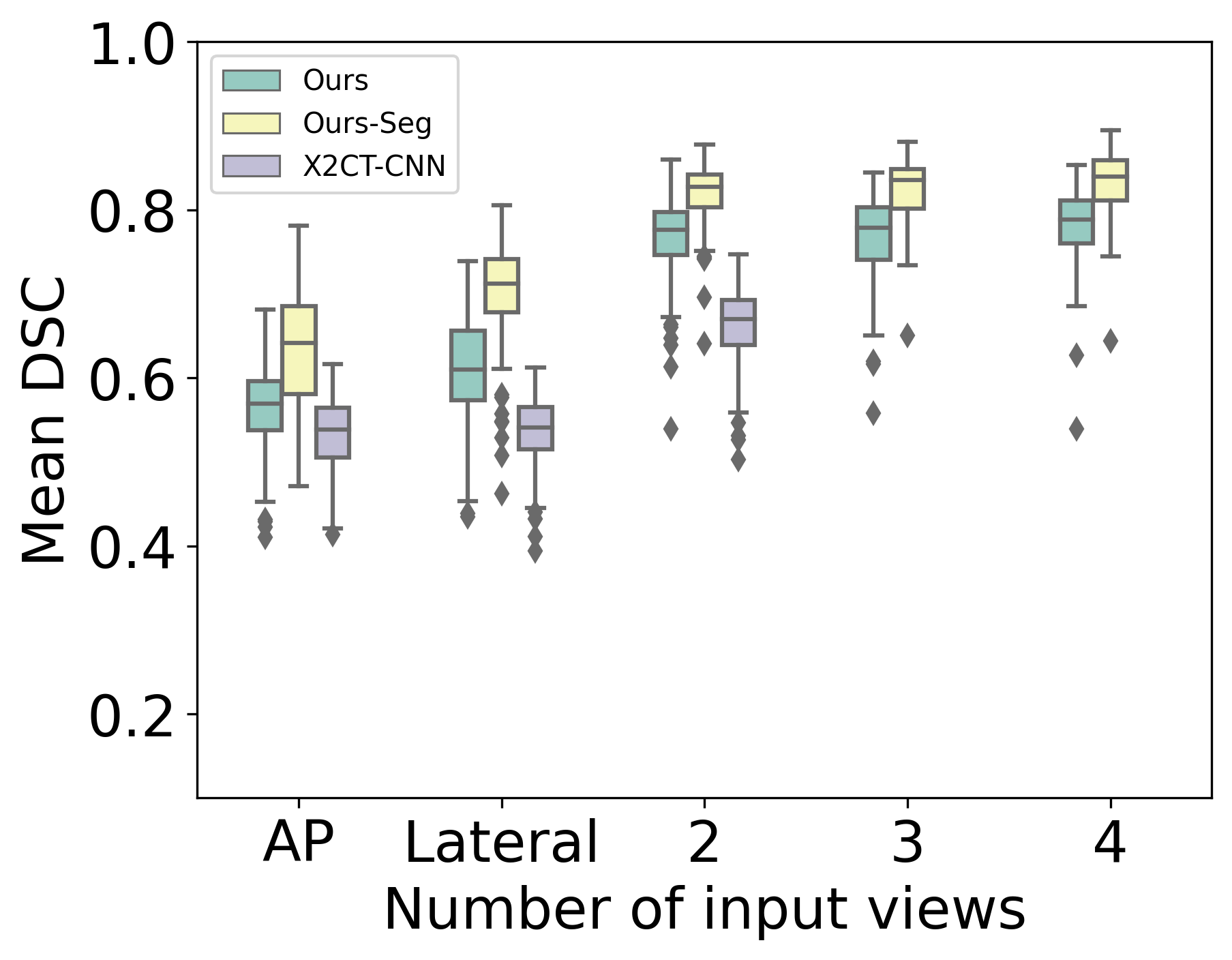}
	\end{minipage}
 }
	\caption{\textbf{Boxplots of PSNR, SSIM, and DSC between reconstructed and ground truth CT scans using 100 test patients in LIDC.} For 1 and 2 views, we also show performance of the baseline X2CT-CNN~\cite{ying2019x2ct} (X2CT-CNN does not work with $>2$ views). Higher values are better.}
	\label{fig:lidc}
\end{figure}

\begin{figure}[t!]
    \centering
    \includegraphics[width=\textwidth]{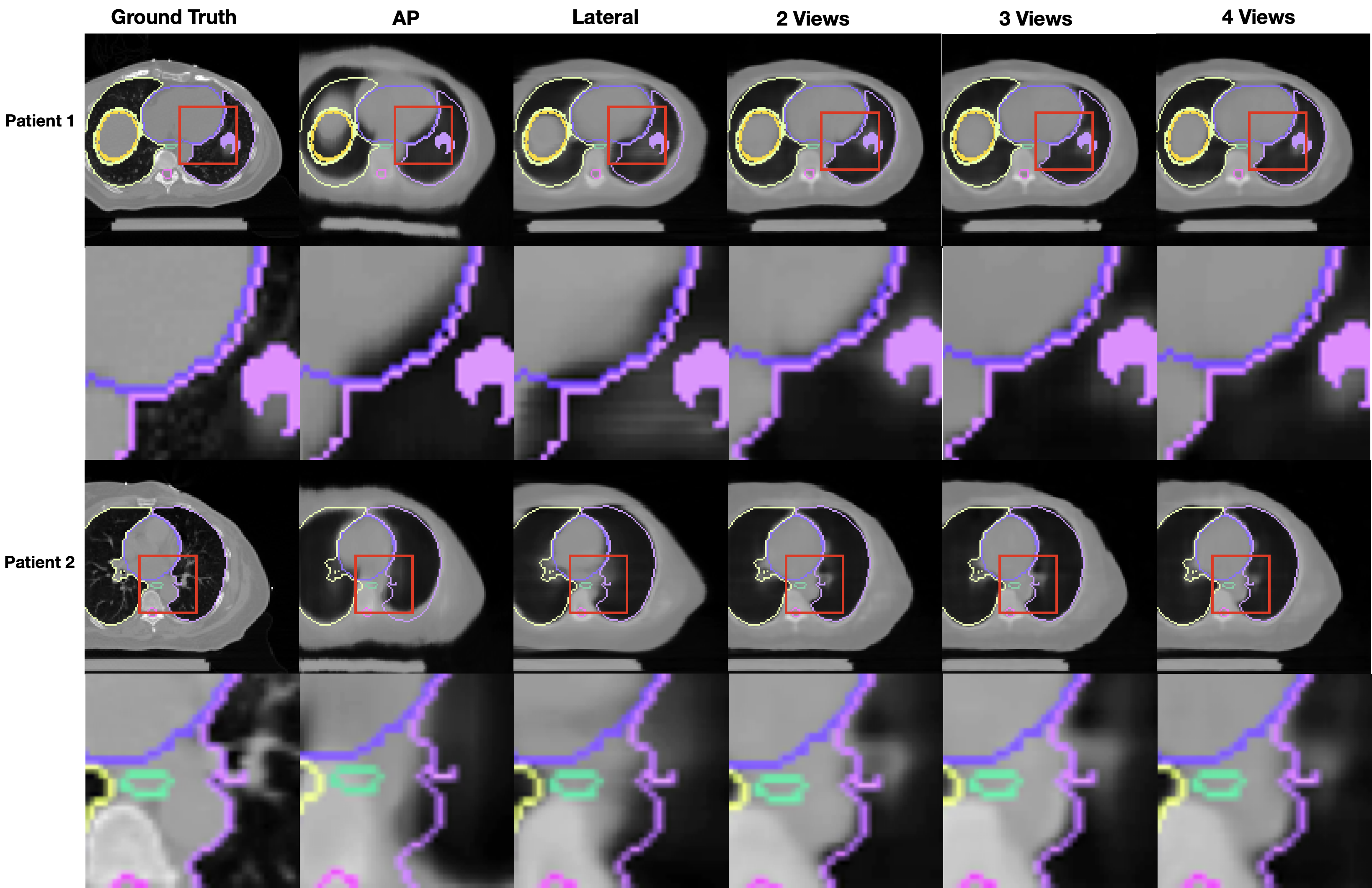}
    \caption{\textbf{Example reconstruction results on Thoracic dataset.} We show results from two patients using the proposed model Ours-Seg. The first column shows the ground truth center slice. The remaining columns show the model's reconstructions for different numbers of input views. The pink contour segments the left lung, and the purple contour segments the heart.}
    \label{fig:CTs}
\end{figure}

\subsection{Results}
First, we compare our models to baselines on LIDC using PSNR, SSIM, and DSC. NAF~\cite{zha2022naf} performs poorly with a few number of views. For example, with 4 input views, 95\% confidence intervals of PSNR, SSIM, and DSC are $23.23 \pm0.81$,  $0.440\pm0.057$, and $0.44\pm0.05$, respectively. Fig.~\ref{fig:lidc} shows the performance of our models and X2CT-CNN. Our models outperform X2CT-CNN for all views, with a particularly striking difference in DSC. 

Moving from 1 to 2 views yields the largest marginal gains. \emph{Ours-Seg} has higher DSC than \emph{Ours}, but has slightly lower PSNR/SSIM. See Supplementary for a table with detailed results. 
\begin{figure}[t!]
\centering
\includegraphics[width=\textwidth, center]{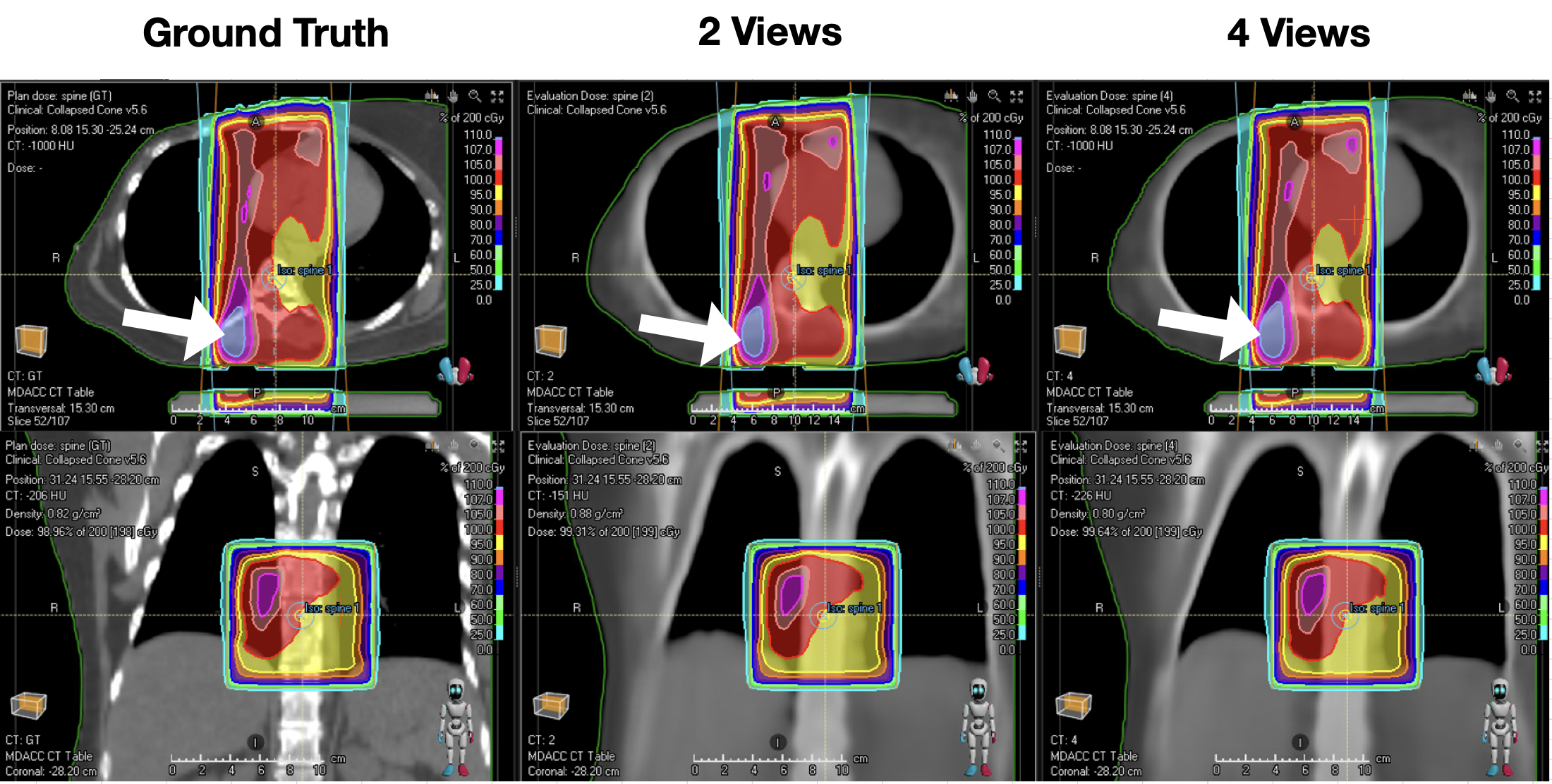}
\caption{\textbf{Radiation planning visual results.} Comparison of radiation plans between ground truth and reconstructed (columns 2-3) CTs for one patient using Ours-Seg. The top and bottom rows represent axial and coronal views of the given CT and overlaid dose distributions. Arrows in the top row indicate the (blue) region of maximum dose. Isodose lines closely match across all scans.}
\label{fig:radiotherapy}
\end{figure}

Next, Fig.~\ref{fig:breast-dsc} presents DSC boxplots of \emph{Ours-Seg} with segmentation training on Thoracic. Again, the largest improvement occurs moving from 1 to 2 views, and the most difficult structures to contour are BrachialPlexus and Esophagus, likely because these structures are small in size. We also show sample reconstruction results with overlaid contours for two patients in Fig.~\ref{fig:CTs} visually confirming the performance improvement near boundaries with more viewing angles.

\begin{table}[t!]
\setlength{\tabcolsep}{12pt}
\caption{\textbf{Average \% errors of isocenter dose on 10 random subjects from Thoracic dataset.} Our models obtain average errors under 1\%. Standard deviations in parentheses.}
		\centering
            \begin{tabular}{clcc}
            \multicolumn{2}{c}{} &
            \cr
            \multicolumn{2}{c}{}         & 2 Views & 4 Views\\ \hline
            \multicolumn{2}{c}{Ours} & 0.30 (0.35) & 0.25 (0.26)\\
            \multicolumn{2}{c}{Ours-Seg} & 0.25 (0.26) & 0.50 (0.57)\\ \hline
            \end{tabular}%
	
\label{tbl:radiotherapy}
\end{table}

Finally, Fig.~\ref{fig:radiotherapy} shows dose distribution results from treatment plans generated on reconstructed CTs using 2 and 4 input views. The shapes of the isodose lines closely resemble that of the ground truth, and in particular, for the high dose region (pointed to by white arrow). Additionally, average isocenter dose errors are under 1\% (see Table~\ref{tbl:radiotherapy}), below the criterion for dose verification accuracy~\cite{zhu2021report}.

\section{Discussion and Conclusion}
Results demonstrate the feasibility of reconstructing CTs from few planar X-ray images. Segmentation guidance during training improves DSC (see Fig.~\ref{fig:lidc}), but did not have a consistent effect on isocenter dose error. The simple planning technique used in this work is used to treat regions in the spine and provide robustness against small uncertainties in patient position. Thus, the radiotherapy dose for this technique is not sensitive to small changes in CT voxel information, which may explain why segmentation-guided training had minimal effects. Further studies using complex, segmentation-driven treatment planning for multiple regions in the body would elucidate the relationship between subtle feature changes in the CT and its impact upon dose.  Structural and dose level metrics presented here indicate that our approach also has potential for use with more complex treatments.

Results also show that our model combining a 2D CNN and INR is better for this task (in terms of voxel level metrics) than an INR only (NAF) that does not leverage prior training data, or a traditional CNN (X2CT-CNN) which suffers in modeling high-frequency details. Maximum performance gain occurs moving from 1 to 2 views, which makes sense since two orthogonal views are generally needed to confirm an object's location within the body~\cite{BRODER2011185}. 

Virtually all existing sparse CT reconstruction studies evaluate results using voxel-level metrics like PSNR and SSIM. This work makes the contribution of additionally evaluating in terms of radiotherapy plans. This is important, because all details need \emph{not} be recovered for an algorithm to be clinically useful, a fact overlooked by PSNR and SSIM. 

There are several exciting next steps to push this work forwards. First, CT reconstruction from few planar X-ray images is a highly ill-posed task and so there are infinitely many possible solutions per planar X-ray image input(s). By returning only one solution, our model is forced to produce scans that are the perceptual ``average'' of possible solutions. Incorporating a probabilistic formulation will help produce sharper results and quantify reconstruction uncertainties. Second, while the inference power of neural networks are remarkable, they are also known to ``hallucinate'' details. We will need further analysis into when and why such models make errors, with a particular focus on atypical subject cases. A focus of our next work will be 3-dimensional conformal radiotherapy planning for chest wall (post-mastectomy). For this specific use case, the tumor and diseased breast tissue is removed before the patient receives a CT and the target of the radiotherapy is the chest wall. Thus, tumor hallucination can be avoided and the development of such a technique will be extremely beneficial for women in low-resource settings, since post-masectomy radiotherapy is extremely prevalent. Finally, in practice we cannot assume that planar X-ray images are acquired at precise angles and depths. Development (and evaluation) of a model that can handle variable acquisition settings would therefore be a valuable contribution.

\section*{Acknowledgment}

This work was supported by NSF CAREER: IIS-1652633.

\bibliographystyle{splncs04}
\bibliography{bibliography}

\end{document}